# Calculation of small arsenic and antimony chalcogenide clusters with an application to vitreous chalcogenide structure

**V. Gurin**

*Research Institute for Physical Chemical Problems, Belarusian State University Leningradskaya 14, 220030 Minsk, Belarus*

**O. Shpotyuk, V. Boyko**

*Institute of Materials of Scientific Research Company "Carat", Stryjska str., 202, 79031 Lviv, Ukraine*

**Abstract**

Small clusters of the As/Sb-S/Se system that is of importance for simulation of elementary structure units of chalcogenide glasses are calculated using DFT technique. Different structures of $As_2X_n$- and $Sb_2X_n$- (X=S,Se) with proper hydrogen termination are compared by the total electronic energy values. The most stable As-X structures are of corner-sharing (CS) type (i.e. the elementary $AsX_3$-pyramids linked via one X atom, and in the case of Sb-X family a new asymmetrical $Sb_2Se_3$ cluster appears rather than its compact ($C_{3h}$) form.

## 1. Introduction

The glasses of the As-X system (X=S, Se) are most extensively studied among different types of vitreous chalcogenide semiconductors [1]. Wide-range elemental composition under preservation of glass-forming ability provides a way to control their physical-chemical properties. The Sb-X glasses are closest analogs of As-X ones, they being of interest from various opinions [2]. A detailed structure of As/Sb-X glasses is a subject of many debates to date [3-5]. One approach to study their structure is a cluster analysis based on construction of molecular clusters built of elementary structural units, such as $AsX_3$, the trigonal pyramidal blocks, those are the small fragments of any glass of this type. In the present work, we consider a series of clusters of $As_2/Sb_2X_n$ composition built by the different ways of joining of the elementary pyramids. We perform DFT quantum chemical calculations to reveal the geometry, electronic structure and relative stability of these clusters, and find their stable structural configurations, which can appear as most frequent glass-forming network elements.

## 2. Model building

The simplest structural element related to As-X glasses is taken as anions of thioarsenite (selenoarsenite) $AsX_3^{3-}$. These anions are rich in many crystal structures like $Ag_3AsS_3$ that allow us to use available crystallographic data for comparison. The reliability of this approach is reflected in data of Table 1.

To calculate the species with one As(Sb) atom as individual neutral clusters (molecules), additional H-termination need, then the composition becomes $AsX_3H_3$. The doubled clusters (-As-X-As-) can be derived from the $AsX_3$-pyramids by joining one with another through three ways: (i) face-sharing (FS), $2AsX_3 \rightarrow As_2X_3$; (ii) edge-sharing (ES), $2AsX_3 \rightarrow As_2X_4^{2-}$, in which the hydrogen termination gives the composition of $As_2X_4H_2$; and (iii) corner-sharing (CS), $2AsX_3 \rightarrow As_2X_5^{4-}$, and the termination results in $As_2X_5H_4$. Thereby this set covers major possible structures of this type for $As_2$-series and will be analyzed below. Fig. 1 displays geometries obtained for the optimized structures and show that close-to-tetrahedron $AsX_3$- units are significantly distorted that may correspond to the flexibility of the glass-forming units.

## 3. Calculation detail

The calculations were done by the DFT method with hybrid functional PBE0 using LANL2DZ basis sets with effective core potential (ECP) for heavy atoms (As,Sb,S,Se) and addition of d-polarization functions (LANL2DZdp); for H atoms the standard 6-311G basis set was taken. The combination of basis/functional has been balanced and tested for best fit with experimental data available (diatomics, crystallography). The NWCHEM package was utilized for geometry optimization and evaluation of electronic structure data [6].

## 4. Results and discussion

Table 1 gives the calculation results for elementary $AsX_3$ units with H-termination, i.e. the $AsX_3H_3$ molecular clusters, served also for verification of geometry data by comparison with similar species in a crystallographic environment, as far as the $AsX_3^{3-}$ anions are quite stable in solids. Good correspondence is evident, that supports adequate usage of the above combination functional/basis for analysis of the clusters with more complicated structures. The calculation data for principal structural parameters of the As-X system (Table 2) demonstrate that for stable As-X structures the distances are rather variable, and the angles are strongly deviate from the values for a perfect pyramid. Binding energies allow determine the relative stability of the clusters treated here in terms of the clusters binding energy with respect to their decomposition on elemental species. The sequence of the structures with decreasing $E_b$ and, correspondingly, $E_b/n_{As\text{-}X}$ appears as CS > ES > FS both for sulfides and selenides. Effective charges given in Fig. 1 evidence weakly ionic character for all As-X covalent bonds and even for $As_2S_3$ the bond polarity is 0. Thus, the clusters of CS type can be considered as the best candidates for models of As-X glass structure. That is in good consistence with previous calculations in other approximations [9] and knowledge of As-X structural features based on a number of experimental data [1,10].

Table 1. Calculation data for the clusters corresponding to elementary building units of the the As-S(Se) system and comparison with crystallographic data

| Cluster, symmetry | $E_{total}$, a.u. | $R_{As\text{-}X}$, Å | $R_{X\text{-}H}$, Å | Bond angles X-As-X, deg |
|---|---|---|---|---|
| $AsS_3H_3$ $C_{3v}$ | -38.3487 | 2.272 | 1.357 | 98.71 |
| $AsS_3^{3-}$ ion in $Ag_3AsS_3$ crystal [7] | | 2.220<br>2.271 | | 101.1<br>95.50 |
| $AsSe_3H_3$ $C_{3v}$ | -35.6936 | 2.415 | 1.489 | 99.51 |
| $AsSe_3^{3-}$ ion in $Ag_3AsSe_3$ crystal [8] | | 2.413 | | 98.53 |

Table 3 presents similar data for a series of antimony selenide clusters. The symmetrical FS structure appears to be very strenuous (Se-Sb-Se angle < $90^o$) while asymmetrical ($C_1$) is more stable instead. However, among these three clusters, the binding energy is maximal for CS-type structure.

## 5. Conclusion

A series of cluster structures of the As/Sb-S/Se system has been calculated using DFT technique with PBE0 functional and ECP basis set. Among the clusters of $As_2(Sb_2)$-X series built from the two elementary pyramids, the CS structures ($C_{2v}$ symmetry) have occurred more stable in the terms of cluster binding energies. They being considered as basic units of further simulations of the chalcogenide glass.

## Acknowledgments

The authors acknowledge support from BRFFR (Project F13K-106) and State Found for Fundamental Research of Ukraine.

Table 2. Calculation data for the clusters of $As_2(S/Se)_yH_z$. $E_b$ is defined with respect to the decomposition of the clusters onto atomic As, X, and molecular $H_2$; $n_{As-X}$ is the number of As-X bonds equal to 6 for all these entities (Fig. 1).

| Cluster symmetry | $E_{total}$ a.u. | $E_b$, eV | $E_b/n_{As-X}$, eV | $R_{As-X}$, Å | $R_{X-H}$, Å | Bond angles X-As-X, deg |
|---|---|---|---|---|---|---|
| **$As_2S_5H_4$ (CS) $C_{2v}$** | **-65.3764** | **32.8137** | **5.5** | **2.275 2.265** | **1.356** | **107.10 102.34** |
| $As_2S_4H_2$ (ES) $C_{2v}$ | -53.9004 | 23.4622 | 3.9 | 2.303 2.525 | 1.361 | 137.55 48.72 |
| $As_2S_3$ (FS) $C_{3h}$ | -42.6531 | 20.3538 | 3.4 | 2.313 | - | 88.60 |
| | | | | | | |
| **$As_2Se_5H_4$ (CS) $C_{2v}$** | **-60.9701** | **29.4722** | **4.9** | **2.4152 2.4078** | **1.486** | **100.77 103.30** |
| $As_2Se_4H_2$ (ES) $C_{2v}$ | -50.4008 | 21.4912 | 3.6 | 2.434 2.638 | 1.489 | 131.80 52.87 |
| $As_2Se_3$ (FS) $C_{3h}$ | -39.6366 | 8.2068 | 1.4 | 2.375 | - | 75.20 |

Table 3. Calculation data for the clusters of $Sb_2Se_yH_z$. $E_b$ is defined similar to Table 2.

| Cluster symmetry | $E_{total}$ a.u. | $E_b$, eV | $E_b/n_{Sb-X}$, eV | $R_{Sb-X}$, Å | $R_{X-H}$, Å | Bond angles X-Sb-X, deg |
|---|---|---|---|---|---|---|
| **$Sb_2Se_5H_4$ (CS) $C_{2v}$** | **-59.5725** | **28.9552** | **4.8** | **2.601 2.590** | **1.488 1.492** | **139.0 139.39** |
| $Sb_2Se_4H_2$ (ES) $C_{2v}$ | -48.9969 | 20.8028 | 3.5 | 2.839 2.638 | 1.489 | 139.92 49.19 |
| $Sb_2Se_3$ (FS) $C_{3h}$ | -36.8583 | -29.8804 | -5.0 | 2.559 | - | 74.3 |
| $Sb_2Se_3$ $C_1$ | -38.6150 | 17.9212 | 3.6 | 2.458-2.653 | - | 99.63 107.61 |

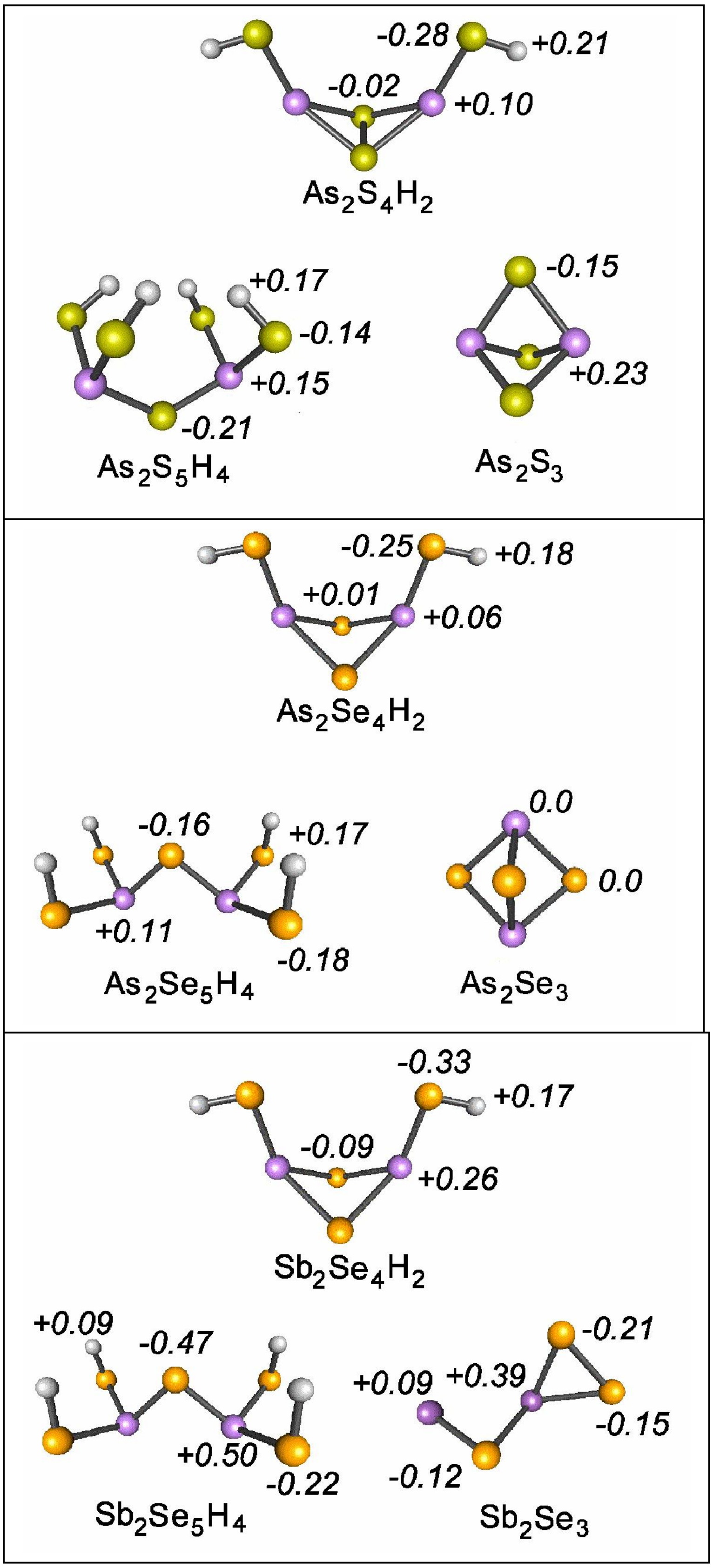

Fig. 1. Optimized geometry of the clusters and effective charges derived from the Mulliken occupancies at symmetry unique atoms.